\documentclass[aps,floats,floatfix,amssymb,amsmath,prb,twocolumn,superscriptaddress]{revtex4-2}
\usepackage[english]{babel}
\usepackage{amsmath}
\usepackage{bm}
\usepackage{graphicx,bbm} 
\usepackage{times}
\usepackage{epsfig} 
\usepackage{soul}
\usepackage[utf8]{inputenc}
\usepackage[colorlinks,linkcolor=blue,citecolor=blue,urlcolor=blue]{hyperref}
\usepackage{color}
\usepackage{latexsym}
\usepackage{ulem}
\definecolor{nred} {RGB}{224,0,0}
\definecolor{nblue} {RGB}{28,130,185}
\definecolor{pgreen}{RGB}{78,138,21}
\definecolor{norange}{RGB}{230,120,20}

\begin{document}

\title{The anisotropic Heisenberg model close to the Ising limit: triangular lattice vs. effective models}
\author{M. Ulaga}
\affiliation{\it Max Planck Institute for the Physics of Complex Systems, Dresden, Germany}
\author{J.  Kokalj}
\affiliation{Faculty of Civil and Geodetic Engineering, University of Ljubljana, SI-1000 Ljubljana, Slovenia}
\affiliation{Jo\v zef Stefan Institute, SI-1000 Ljubljana, Slovenia}
\author{T. Tohyama}
\affiliation{Department of Applied Physics, Tokyo University of Science, Tokyo 125-8585, Japan}
\author{P. Prelov\v{s}ek}
\affiliation{Jo\v zef Stefan Institute, SI-1000 Ljubljana, Slovenia}

\begin{abstract}
  Stimulated by recent experiments on materials representing the
  realization of the anisotropic Heisenberg spin-$1/2$ model on
  triangular lattice, we explore further properties of such a model in
  the easy-axis regime $\alpha = J_\perp/J_z < 1$ and the plausibility of
  finding effective models that capture similar physics.  We show that, at finite fields, 
  the magnetization curve as well as the transverse
  magnetization (superfluid) order parameter $m_\perp$ of the triangular lattice model are indeed qualitatively reproduced by anisotropic Heisenberg models on the honeycomb or the square lattice. At the point of correspondence to the zero-field triangular lattice model, however, the bipartite models are qualitatively different as they remain gapless even at $\alpha \ll 1$ with a small but finite $m_\perp >0 $. Conversely, we present several
  additional numerical studies of the full model on the triangular lattice
  which support the appearance of a gap at zero field and $\alpha \ll 1$. In particular,
  the magnetization curve $m(h)$ as well the spin stiffness $\rho_s$
  indicate a transition/crossover from gappless to
  gapped regimes at $\alpha \sim \alpha^*$ with
  $\alpha^* \lesssim 0.5$.  We also show that deviations from the
  linear spin-wave theory and the emergence of the gap can be traced
  back to the strong effective repulsion between magnon excitations,
  showcasing similarity to strongly correlated systems.
       
\end{abstract}

\maketitle

\section {Introduction}

The antiferromagnetic Heisenberg spin-$1/2$ model on the
triangular lattice (TL) has been a source of many novel theoretical 
concepts as well as challenges. The isotropic case has been first
proposed as the model for the phenomenon of quantum spin liquid
\cite{anderson73}, but later shown to exhibit in the ground state (gs) 
a nontrivial broken translational symmetry with tripling of the unit cell and 
$120^0$ alignment of spins \cite{bernu94,chernyshev09},
being already a challenge for powerful numerical 
approaches \cite{capriotti99,white07}. The anisotropy $\alpha = J_\perp/J_z$ 
in the easy-axis regime  $|\alpha| < 1$  opens another set of phenomena. 
While the Ising limit with $\alpha =0$ has been solved analytically
to reveal a finite entropy at $T=0$ \cite{wannier50}, the regime 
$0 < |\alpha| < 1$  considered as an interesting problem long ago  
\cite{miyashita85} has emerged as the potential realization of the
scenario of (spin) supersolid \cite{boninsegni12}, having (at least at $T=0$), 
besides the broken translational symmetry  and related diagonal magnetization 
$m_z > 0$, also  broken rotational symmetry with finite off-diagonal transverse 
magnetization $m_\perp  > 0$. This possibility has been studied and established
in the regime $\alpha < 0$ by a number of numerical studies   
\cite{heidarian05,boninsegni05,wessel05,wang09},
with apparently similar results for the antiferromagnetic 
regime $\alpha > 0$  \cite{wang09,jiang09,yamamoto14,sellmann15}.

The strong additional momentum to theoretical considerations has
recently been given by the synthesis and experiments on novel materials 
where the properties appear to be well represented by the antiferromagnetic
anisotropic Heisenberg model (AHM) on TL with $0< \alpha <1$.   
While Na$_2$BaCo(PO$_4$)$_2$ \cite{li20,gao22,xiang24,gao24} with $\alpha \sim 0.6$ 
seems to represent  the spin system with properties close to the isotropic case  
and NdTa$_7$O$  _{19}$ \cite{arh22} is characterized by  $\alpha \ll 1$ but small $J_z\sim 0.9\,\textrm{K}$, 
the most promising material is  K$_2$Co(SeO$_3$)$_2$ (KCSO) \cite{zhong20}
with $\alpha \sim 0.07$ and $J_z\sim 35\,$K, which has already allowed experimental studies of thermodynamic properties,
as well as dynamical spin excitations (via inelastic neutron scattering) in a wide range of 
temperatures and magnetic fields \cite{zhu24,chen24,zhu124}.  
These experimental findings have already stimulated a series of additional (mostly numerical) 
theoretical  studies  \cite{ulaga24,ulaga25,gallegos25,xu25,flores25}. 

While in many aspects these theoretical (model) calculations agree with experimental findings, the 
potential qualitative disagreement is the strict existence of the supersolid, i.e., of 
gapless excitations (and corresponding $m_\perp >0$) in the AHM  at $h=0$ 
(without external magnetic field)  close to the Ising limit, i.e., at $\alpha \ll1$.  In particular,
the available numerical results for the supersolid parameter $m_\perp$ at $\alpha \ll1 $ 
indicate at least its very small value, being strongly dependent 
on applied finite-size scaling $N \to \infty$ \cite{jiang09,ulaga24,ulaga25,xu25,gallegos25}. This might 
indicate even a vanishing  $m_\perp(h=0) = 0 $ \cite{ulaga25,xu25}, compatible with the interpretation with
gapped magnon excitations $\Delta_1 > 0$ and discontinuous $T=0$ magnetization 
curve $m(h < h^*) =0$  \cite{ulaga25}, with $h^* \ll h_c$ where $h_c$ denotes the onset
of the  $m = 1/3$ magnetization plateau well pronounced and extended in the $\alpha \ll 1$ regime. 
On the other hand, numerical studies 
as well as the experiment are consistent in the observation that $h^*< h < h_c$ induces supersolid 
$m_\perp >0$. 
   
Since the AHM on TL in the $\alpha \ll 1$ regime  at $T=0$ reveals the tripling of the unit cell
and large longitudinal magnetization $m_z > 0$,  one might speculate that some properties of the full 
model can be reproduced within a reduced model, whereby breaking the translational
symmetry, a third of spins are 
fixed in their maximum value, leading to an effective AHM on the honeycomb lattice (HcL), but 
in a finite field $h>0$ \cite{ulaga25}, see Fig.~\ref{fig1a} for an illustration. Since the latter lattice is (in contrast to TL) bipartite, even 
AHM on a square lattice (SqL) could be expected to have similar properties.  We present here
a detailed $T=0$ numerical study of the corresponding magnetization curve $m(h)$ as well as 
$m_\perp$, which reveals a correspondence to the full TL regarding the anomalous behavior 
at $\alpha  \ll 1$ and $h>0$. 

Still, despite this similarity, one clear message from studying HcL and SqL is that at the most interesting point of correspondence (to the TL with $h=0$), a qualitative difference remains: magnon excitations remain gapless on these bipartite models and they possess a finite $m_\perp >0$, although strongly reduced. Returning to the most challenging problem, i.e., the AHM on TL and $h \sim 0$, we 
present additional evidence that in the regime of $\alpha \ll 1$ we are dealing instead with a 
gapped solid, with the support emerging from numerical results for magnetization curves 
$m(h  \sim 0)$ and for the spin stiffness $\rho_s$. Such a scenario naturally opens the 
question of the presumable disappearance of the gap and the onset of $m_\perp > 0$ with increasing 
$\alpha$, in particular on approaching the isotropic gapless case $\alpha = 1$. Consistent with our previous limited evidence \cite{ulaga24}, our additional results indicate a 
qualitative change, i.e., a crossover/transition at $\alpha^* \lesssim 
0.5$, which evidently remains a challenge for a proper phenomenological explanation
and further numerical tests.  On the other hand,  the existence of gapped excitations at $0 < \alpha \ll 1$ also reopens the question on the relation to the regime $\alpha < 0$ where the supersolid scenario appears more
firmly established \cite{heidarian05,boninsegni05,wessel05,melko05,wang09}. We therefore extend 
our analysis to this regime, which, despite apparent symmetry, also reveals qualitative
differences.       

Since theoretical and experimental results are frequently compared to the 
linear spin-wave theory (LSWT)  \cite{toth15}, we also critically evaluate its feasibility to 
AHM in the $\alpha \ll 1$ regime. It has already been recognized that for the TL, the latter yields $m_\perp(h)$
monotonously increasing with decreasing $h<h_c$ \cite{ulaga25} with evident discrepancy  with 
numerically established  vanishing \cite{ulaga25,xu25} (or at least very small \cite{gallegos25})
$m_\perp(h = 0)$,  whereby  the origin could be attributed to the strong repulsion/correlations 
between magnons \cite{kleine92,mauri25}, which still represents a challenge for
adequate analytical analysis.

 \section{Reference models on honeycomb and square lattice}
 
We consider the antiferromagnetic $S=1/2$ AHM with the nearest-neighbor (nn) exchange 
$J_z =J$ and the easy-axis anisotropy $0 < \alpha \le 1$, in the presence of a longitudinal  
magnetic field $h$,
\begin{equation}
H= J  \sum_{\langle ij \rangle} \lbrack S^z_i S^z_j + \frac{\alpha}{2}( S^+_i S^-_j  + S^-_i S^+_j)  \rbrack
- h\sum_{i}  S^z_i~~,~~\label{his}
\end{equation}
on different lattices with nn interaction and $N$ sites.  While the
main challenge remains the model on TL, we first focus on the  reduced
model on HcL. This emerges naturally as the
representation of the full TL model by taking into account that at $\alpha \ll 1$
there is broken translational symmetry (at least at $T=0$) with tripling of the unit cell
and a well-pronounced diagonal long-range order parameter 
$m_z = (1/N) \sum_i \exp(i {\bf q}_K \cdot {\bf R}_i) \langle S^z_i \rangle \neq 0$.
where ${\bf q}_K = (4\pi/3,0)$. 
\begin{figure}[ht]
\centering
\includegraphics[width=0.6\columnwidth]{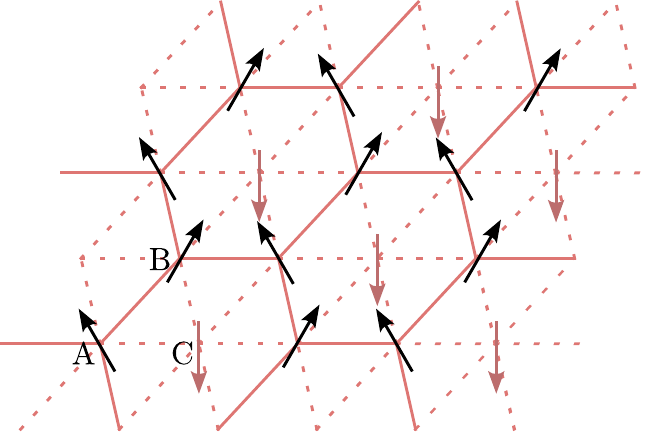}
\caption{An effective Heisenberg spin model on the honeycomb lattice (HcL), 
where the translational symmetry on TL is 
broken by fixing a third of spins of the classical Y state on the triangular lattice (sublattice C). The remaining two sublattices (AB) form a magnetized HcL under the effective field of the frozen spins.}
\label{fig1a}
\end{figure}
Such a situation can be simulated by freezing/fixing one sublattice of spins on TL
to their maximum polarization $S_i = -S$ \cite{ulaga25} as illustrated 
in Fig.~\ref{fig1a} (see also 
Ref.\cite{mauri25}), which leads to the model
Eq.~(\ref{his}) on HcL, but at an
additional effective field $\tilde h = z J S$, 
where $z=3$ is a number of nn on HcL.  It should be noted
that the correspondence of HcL to TL model becomes exact (at
$\alpha \to 0$) at the magnetization plateau $ m= S^z_{\textrm{tot}}/ (NS) = 1/3$
since there the $\uparrow \uparrow \downarrow$ spin pattern is the
ground state (gs) of the model on TL, and matches with the
ferromagnetic solution $S^z_{\textrm{tot}} = S N$ and corresponding
$ m = S^z_{\textrm{tot}} / (NS) = 1$ on HcL.  Starting from the $m=1/3$
plateau, the qualitative correspondence persists for finite
$\alpha > 0$, while the central question is to what extent the
effective HcL model reflects the physics of the  TL when the HcL
magnetization is decreased to $ m = 1/2$ \cite{ulaga25}, simulating
$m=h=0$ in the TL model.  The HcL is a bipartite lattice with two
sites per unit cell. This naturally opens the possible similarity of
the AHM on an even simpler bipartite lattice, i.e., AHM on SqL 
\cite{holtschneider07}, where $z=4$ and the focus is on
the properties at effective $h > 0$, or equivalently on magnetizations
$1/2 \lesssim m < 1$.

\subsection {Magnetization curves}  

\begin{figure}[t]
\centering
\includegraphics[width=\columnwidth]{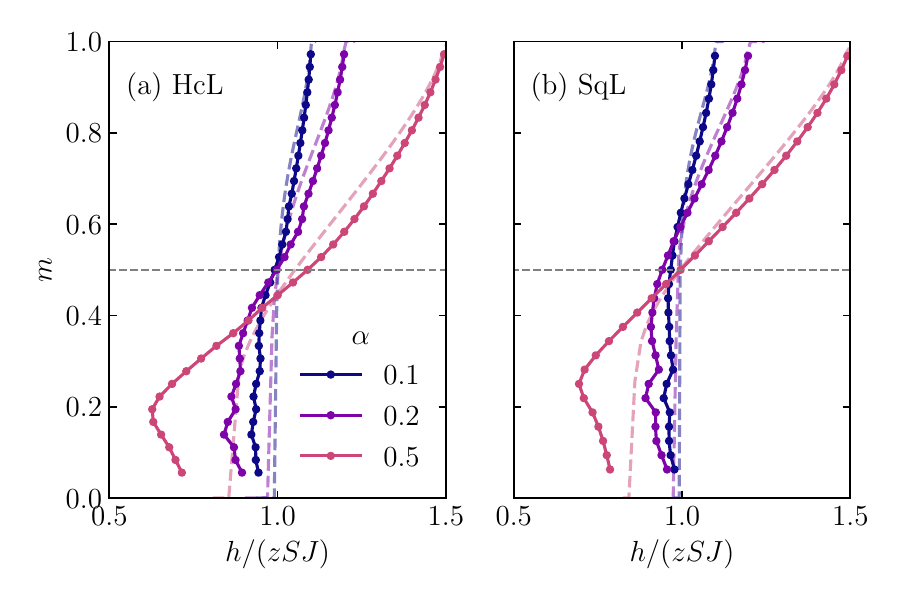}
\caption{Magnetization curves $m(h)$ vs. renormalized fields $\tilde h = h/(zSJ)$ for (a) 
the honeycomb lattice (HcL) and ( b) the square lattice (SqL),  for different anisotropies 
$\alpha = 0.1, 0.2, 0.5$, obtained with DMRG calculation on lattices 
with $N=72$ sites and $N=64$ sites, respectively. The black dashed
lines indicate $m=1/2$ and $\tilde h = 1$,  
corresponding  to $m=0$ and $h=0$, respectively,  within the TL model.
The coloured dashed lines represent the augmented LSWT results for $m(\tilde h)$, discussed 
in Sec.~IV.}  \label{fig1}
\end{figure}

Within the effective models on HcL and SqL we first consider and
evaluate magnetization curves $m(h)$. As in the previous study of
AHM on TL \cite{ulaga25}, the DMRG calculation on finite-size lattices
with $N$ sites and with periodic boundary conditions (PBC) is here
employed to find gs energies $E^0_k$ in different magnetization
sectors $k = S^z_{\textrm{tot}}$.  Actual results in Fig.~\ref{fig1}a,b are
obtained on HcL with rhombic cluster of $N_c = 6 \times 6$ unit cells
(i.e., with $N=2N_c = 72$ sites), and on SqL with a square
$N = 8 \times 8 =64 $ cluster (see Appendix~\ref{dmrg}).
From $E^0_k$ and magnetizations
$m_k = k /(NS) $, the corresponding fields are then extracted as
$h_k = (E^0_{k+1} - E^0_{k-1})/2$.

Results on Fig.~\ref{fig1}a,b show $m(\tilde h)$
vs. renormalized fields $\tilde h = h/(zSJ)$ on HcL (with the nn
number $z=3$) and SqL ($z=4$), respectively, for different
$\alpha = 0.1, 0.2, 0.5$. The first observation is that, indeed, the results are
qualitatively and even quantitatively similar with respect to
the variation with $\alpha$, taking into account different $z$.  We note
that in the Ising limit $\alpha =0$ the AFM solution becomes unstable
to ferrromagnetic one at $\tilde h = 1$ \cite{honecker99}.  For the
isotropic case with $\alpha = 1$ (not shown here), it has been
established that for SqL $m(h)$ is monotonously increasing from $h=0$
to $h=h_c = 2J$ \cite{honecker04}, and quite similar behavior is
expected for HcL.  For $\alpha < 1$, there is a qualitative change at
$h \to 0$, since magnon excitations in the antiferromagnetic gs of AHM
at $h=0$ become gapped with the magnon gap
$\Delta_1 \sim zJ S \sqrt{1-\alpha^2}$ \cite{holtschneider07}.  This
leads to stable $m=0$ solution for $h < h_1$ and for nonmonotonous
(double solution) $m(h)$ in the regime $h_1 < h < h_2$. This
instability regime (indicating a first-order transition of the gs
with $h$) shrinks with decreasing $\alpha \to 0$, ending in a single
step in $m(\tilde h)$ in the Ising limit at
$\tilde h_1=\tilde h_2 = 1$.

Still, our focus is on the behavior at finite $\alpha > 0$ and the
magnetization $m \gtrsim 1/2$, where the correspondence with the TL at
$h \gtrsim 0$ is meaningful.  In contrast to discontinuous
$m(h < h^*)=0$ (consistent with a magnon gap $\Delta_1 > 0$ ) found in
TL \cite{ulaga25}, in Fig.~\ref{fig1} it appears that at $m \sim 1/2$
we have continuous $m(h)$ and $dm/dh > 0$ for both HcL and SqL, i.e.,
there is no indication of a finite magnon gap. Conversely,
the regime $m<1/2$ is quite different on HcL as well as on SqL,
revealing the (first-order) transition to the stable antiferromagnetic
$m=0$ solution, as well visible in Fig.~\ref{fig1} for all presented
$\alpha < 1$.

We should comment here on the previous observation of anomalous
excitations on the HcL at small $\alpha = 0.1$ and $m =1/2$
\cite{ulaga25}.  They were obtained via exact diagonalization (ED) on
smaller $N=40$ as excited states within the same $S^z_{\textrm{tot}}$, thus
effectively representing two-magnon excitations, with $\omega_{\bf q}$
revealing an unusual upturn at the smallest $q$. The latter result is another
manifestation of the delicate behavior of effective models at
$m \sim 1/2$, which is shown here to persist for considerably larger
$N$. The calculated $dm/dh > 0$ are thus a more direct test against a
possible finite magnon gap.

\subsection{Transverse magnetization}

In both effective models, one can expect at $T=0$ and $h>0$ broken
rotational spin symmetry and corresponding order parameter, i.e., the
transverse magnetization $m_\perp > 0 $. We extract $m_\perp$ via the $T=0$ transverse dynamical
spin structure factor (DSSF)
\begin{equation}
S^\perp({\bf q},\omega) =
\langle \psi_0| S^x_{-\bf q} \delta(\omega - H + E_0) S^x_{\bf q}| \psi_0 \rangle,
\end{equation} 
evaluated within the gs wavefunction $|\psi_0 \rangle$ for given spin
sector $S^z_{\textrm{tot}}$.  Here, we calculate DSSF on finite HcL and SqL with
PBC via ED, employing the Lanczos technique (see, e.g.,
\cite{prelovsek13}).  Analyzing the lowest (soft) mode
$\omega_{{\bf q}_0} = \Delta_1$ enables calculating the
corresponding
$S^\perp({\bf q}_0,\omega) \sim A^\perp \delta( \omega - \Delta_1)$
and finally extracting $m^2_\perp = A^\perp/N$.  On HcL the relevant
rotational-symmetry breaking appears at ${\bf q}_0 = (0,0)$ with the (odd
in each unit cell) spin operator
$S^x_0= N^{-1/2} \sum _i ( S^x_{i1} - S^x_{i2}) $, summed over
all unit cells.  Conversely, for the SqL the soft mode is the
zone-boundary one with ${\bf q}_0 = {\bf q}_M =(\pi,\pi)$ and the
corresponding
$S^x_{{\bf q}_M}= N^{-1/2} \sum _i e^{i {\bf q}_M \cdot {\bf R}_i}
S^x_i $.

\begin{figure}[t]
\centering
\includegraphics[width=\columnwidth]{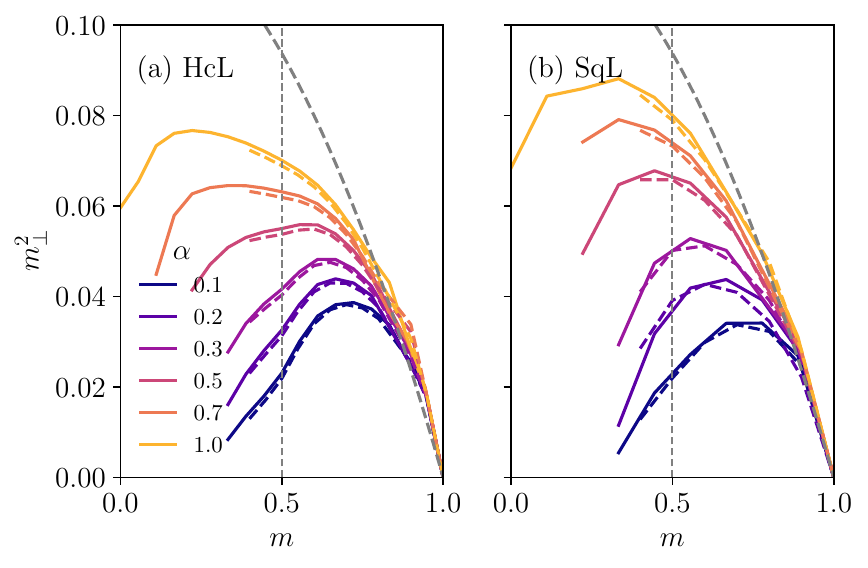}
\caption{Transverse order parameter $m_\perp^2$ vs. magnetization $m$,
extracted from ED results for DSSP on systems with $N =36$ and $N=40$ sites on 
(a) HcL, and (b) SqL, for different $\alpha = 0.1 - 1.0$. Vertical $m=1/2$ line indicates the correspondence 
to the $h=m=0$ model on TL. The gray dashed curve represents the expected LSWT dependence 
$m^2_\perp = \zeta(1 - m^2)$ with $\zeta = 1/8$. }  \label{fig2}
\end{figure}

In Fig.~\ref{fig2}a,b, we present results for $m_\perp^2$
vs. magnetization $m = 2 S^z_{\textrm{tot}}/N$, obtained via ED on $N =36, 40$
sites for HcL and SqL, shown within the whole range of anisotropies,
$\alpha = 0.1 - 1.0$.  The advantage of a smaller $N=36$ is that via ED
we can cover the whole range $0 \leq m \leq 1$, whereas for $N=40$ we are
limited (due to Hilbert-space dimension) to $m> 0.2$. Also, generally
we restrict in Fig.~\ref{fig2}a,b our presentation to
$m > m_*(\alpha)$, since the regime $m < m_*(\alpha)$ corresponds to
the unstable regime in $m(h)$ and results become unphysical (as well as
strongly size dependent).

Again, the qualitative behavior of
$m_\perp^2(m)$ is quite similar for both HcL and SqL. As expected, the
dependence is quite generic, i.e., independent of $\alpha$, for nearly
saturated $m \lesssim 1$. There, one can give an explanation with the
simplest classical result of the LSWT, where $m = \cos(\theta)$ and
$m_\perp = S \sin(\theta)$ (note our different  definitions of $m \leq 1$ and
$m_\perp \leq 1/2$) and consequently
$m^2_\perp \propto 1 - m^2$.   Still, deviations become very
pronounced with decreasing $m < 1$ and $\alpha \ll 1$. At all
$\alpha$ maximum of $m_\perp(m=m_0)$ is for $m=m_0 >0 $, a weak one even
for the isotropic case $\alpha = 1$, with the value
$m_\perp(h=m=0) = 0.26$ for SqL, being
close to the established values 
\cite{manousakis91}.  On the other hand, the maximum becomes very
pronounced for $\alpha < 0.5$, typically appearing at
$m_0 > 0.5$.
 
\begin{figure}[t]
\centering
\includegraphics[width=\columnwidth]{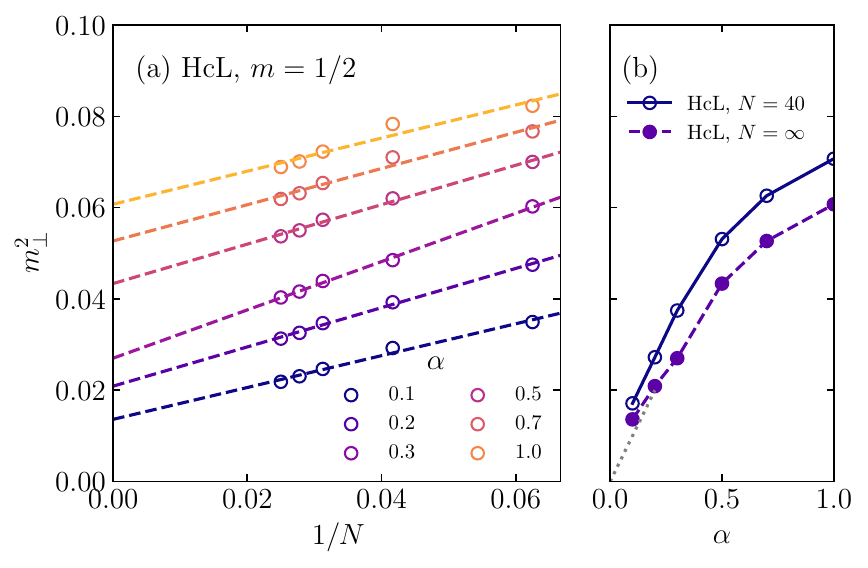}
\caption{ Transverse order parameter $m_\perp^2$ at $m = 1/2$ for HcL : (a) finite-size $1/N$ scaling 
of results for different $\alpha$, (b) $m_\perp^2$, obtained on  $N=40$ sites,
and extrapolated $N \to \infty$ values vs. $\alpha$.}  
\label{fig3}
\end{figure}

Fig.~\ref{fig3}a presents the $1/N$ scaling of $m_\perp^2$ results for
HcL at the most interesting $m =1/2$. In contrast to an analogous analysis
for TL, they reveal rather evident convergence to finite
values even for the smallest $\alpha = 0.1$.  In Fig.~\ref{fig3}b, we show $m_\perp^2$ vs. $\alpha$, both for $N=40$ and
extrapolated $N \to \infty$ values. The similarity to the TL is evident: 
for $\alpha \ll 1$ the pronounced maximum of $m_\perp(m)$ appears at intermediate $0.5 < m < 1$
(in contrast to LSWT prediction). However, a qualitative difference persists in the numerical results for 
$m_\perp(h=0) \sim 0$  on TL for $\alpha \ll 1$, extrapolating to much smaller values
(essentially too small to be reliable) \cite{ulaga25}. It is an open question 
why effective models still behave differently with respect to the gapped/gapless excitations
at the point of correspondence. One evident and possibly crucial difference might emerge
from the fact that the full model on TL at $h=0$  has an inversion symmetry $m \to -m$, broken on HcL and SqL at $m=1/2$.
 
 \section{Triangular lattice: evolution with anisotropy}
 
 We turn back to the most challenging as well as experimentally
 most relevant case of AHM on TL at small $\alpha \ll 1$. In spite of
 similarities with effective models at the point of correspondence
 (i.e., $m \sim 1/2$ for HcL and SqL), there are qualitative
 differences with respect to the existence (or at least the size) of
 the transverse magnetization (supersolid parameter) $m_\perp > 0$ in
 the absence of an external field, at  $h=0$.  As noticed already in several
 numerical studies of the $0 < \alpha \ll 1$ regime on TL
 \cite{jiang09,ulaga25,xu25,gallegos25}, the extracted values for
 $m_\perp $ are very sensitive to the finite-size extrapolation
 $ N \to \infty$.  In the following, we present some additional
 support for the scenario that at $\alpha \ll1 $ we are dealing at
 $h=0$ with the absence of supersolidity, i.e., $m_\perp =0$
 \cite{ulaga25,xu25}, which can be made compatible with the finite
 magnon gap $\Delta_1 > 0$. The most challenging question in this
 respect is the evolution of the behavior with increasing
 $\alpha \to 1$, since at least the gs of the isotropic case $\alpha = 1$
 represents an ordered magnet with $120^0$ spin alignment and 
 gapless magnon excitations \cite{chernyshev09}. 
 
 \subsection{Magnetization curves}
 
\begin{figure}[t]
\centering
\includegraphics[width=0.8\columnwidth]{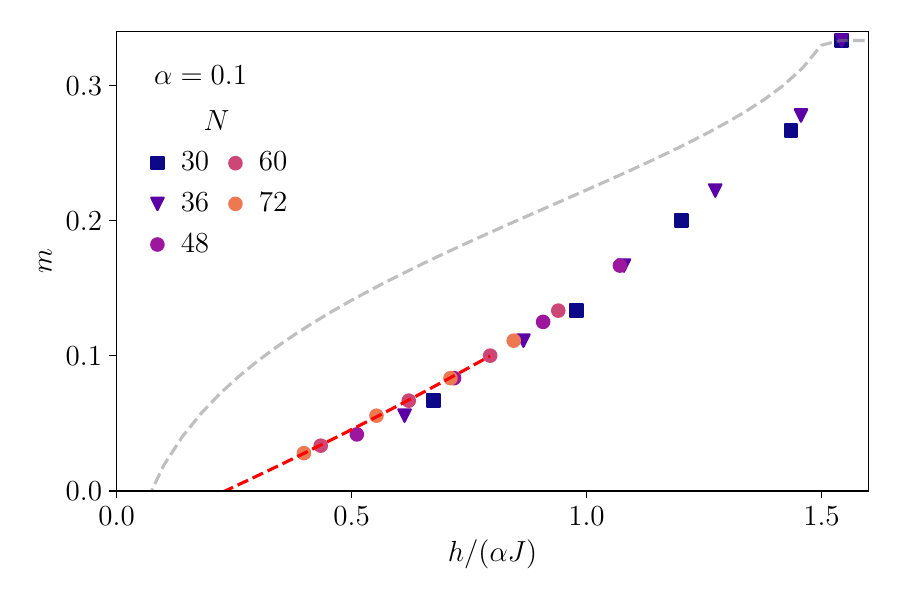}
\caption{ Magnetization curve $m$ vs. normalized field $h/(\alpha J)$, for $\alpha =0.1$, 
obtained with ED for $N = 30, 36$ lattices and via DMRG for $N=48-72$ lattices, with the 
quadratic extrapolation (red dashed line). The black dashed line represents 
the result of LSWT, explained and discussed in Sec.~IV.} 
\label{fig4}
\end{figure}

\begin{figure}[t]
\centering
\includegraphics[width=0.7\columnwidth]{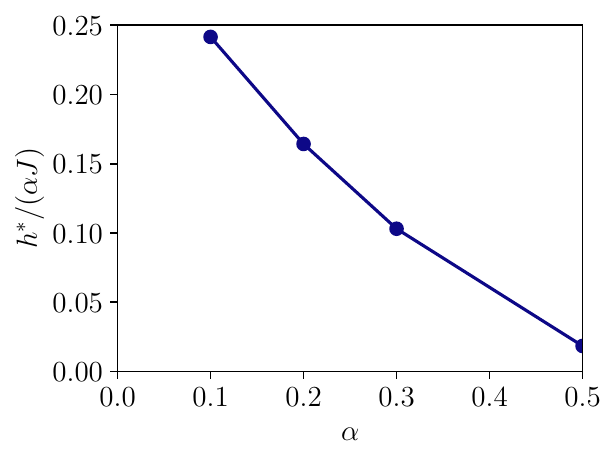}
\caption{ Scaled gap $h^*/(\alpha J)$ vs.  $\alpha$, as obtained by
extrapolation of $m(h)$ results presented in Fig.~\ref{fig4} and in 
Appendix~\ref{chap_append}. } \label{fig5}
\end{figure}

 A sensitive test of the possible gap is the magnetization curve $m(h)$
 which is also a directly experimentally relevant quantity, at least for KCSO
 \cite{zhu124}. A finite magnon gap $\Delta_1$ implies
  vanishing $m(h<h^*=\Delta_1) =0$ at $T=0$.  For particular
 $\alpha =0.1$ the evidence for finite $h^* >0$ has been presented
 with the previous calculations up to $N=60$ model on TL
 \cite{ulaga25}. Here, we extend the DMRG study to larger 
 $N =72$, but also to more systematic evolution with $\alpha$
 including also $\alpha < 0$ for comparison. We present in
 Fig.~\ref{fig4} the results for $m(h)$ for the characteristic value
 $\alpha = 0.1$, obtained with ED for $N \leq 36$ and
 DMRG for $N = 48, 60, 72$ (restricted to $S^z_{\textrm{tot}} \le 4$).
A more systematic analysis of the evolution of $m(h \to 0)$
with anisotropy $\alpha \leq 1$, including also 
$\alpha = -0.1, -0.2$, is delegated to Appendix~\ref{chap_append}.
The presentation in terms of renormalized $h/(\alpha J)$ reveals
 a quite similar dependence, apart from the crucial regime $h \to 0$.
 Regarding the latter, there is clear evidence that the extrapolated results
 would be consistent with quite different marginal fields (effective magnon gaps) 
 $h^* = \Delta_1 = \zeta \alpha J$, as summarized in Fig. \ref{fig5}. Results
 indicate that renormalized $h^*/(\alpha J) = \zeta$ is strongly reduced at
 intermediate $\alpha^* \sim 0.5$, which is an indication of a 
 transition or crossover to a gapless regime. We should also point out a clear asymmetry
 with respect to the sign $\pm \alpha$, since at least $\alpha = -0.2$ (less clear for $\alpha = -0.1$)
 is as expected \cite{heidarian05,boninsegni05,wessel05,wang09} consistent with
 gapless excitations and consequently supersolid order.   
 
 \subsection{Spin stiffness}
 
 Another important measure and test of the gs character is the spin
 stiffness $\rho_s$, representing the sensitivity of the
 supersolid order parameter $m_\perp$ to a twist in boundary conditions \cite{melko05,jiang09}.  In
 finite-size calculations, it can be evaluated by introducing in
 Eq.~(\ref{his}) a phase
 $\theta_{ij} = \theta ~{ \bf e}_\theta \cdot {\bf R}_{ij} $ into the
 spin exchange term,
\begin{equation}
\tilde H_{ex} =  \frac{1}{2} J \alpha \sum_{\langle ij \rangle} [\mathrm{e}^{i \theta_{ij} }S^+_i S^-_j  +
\mathrm{e}^{- i \theta_{ij}} S^-_i S^+_j ].
\end{equation}
The spin stiffness is then calculated as
$\rho_s = (1/N) \partial^2 E_0/ \partial^2 \theta$
\cite{bonca94,lecheminant95,kruger06}. In analogy to the charge stiffness
\cite{kohn64} and the superfluid stiffness \cite{scalapino93}, finite
$\rho_s >0 $ indicates gapless excitations, while $\rho_s =0 $ is the
signature of a gapped system.

\begin{figure}[t]
\centering
\includegraphics[width=0.8\columnwidth]{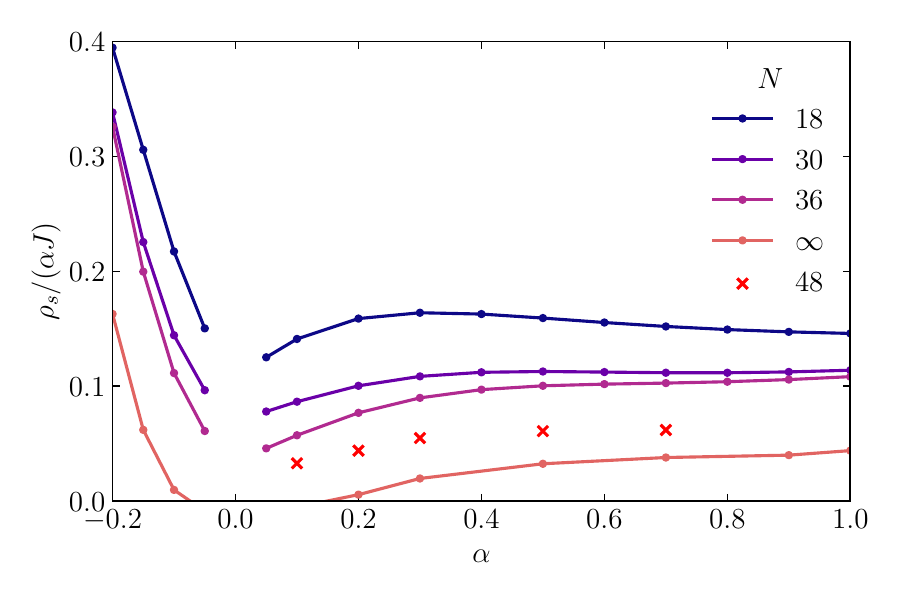}
\caption{ Normalized spin stiffness $\rho_s/(\alpha J)$ vs. anisotropy $\alpha$, as calculated
on TL via ED on systems with different sizes $N = 18, 30, 36$, together with the extrapolated 
value $N \to \infty$, obtained via the $1/N$ scaling.  Crosses denote DMRG results on $N=48$ sites.} \label{fig6}
\end{figure}

We present in Fig.~\ref{fig6} results for renormalized
$\rho_s/(\alpha J)$ vs. $\alpha$, including also the range
$\alpha <0$, obtained within the ED approach by applying small
$\theta \sim 0.1$ in the $x$ direction, as well those obtained 
via DMRG on $N = 48$ sites. The results in Fig.~\ref{fig6}
reveal quite significant finite-size dependence (decrease) of
calculated $\rho_s$ with the system size $N$.  Still, for the isotropic case
$\alpha =1$ the $1/N$ extrapolation yields finite value
$\rho_s \sim 0.05 J$, well consistent with previous results for TL
\cite{kruger06}. For the present study, the most interesting and
challenging aspect is the development of normalized $\rho_s/(\alpha J)$, in
particular its extrapolated value, with $\alpha$.  As evident in
Fig.~\ref{fig6} the variation is quite modest for $\alpha >
0.4$. However, there is a qualitative change at
$\alpha < \alpha^* \sim 0.2$, indicating a vanishing $\rho_s = 0$,
which can be compatible with the onset of finite gap $\Delta_1 > 0$.
In this respect, results in the $\alpha < 0$ regime
are also quite informative: some similarity is visible (e.g., finite-size
sensitivity), but also a pronounced asymmetry with
the regime with $\alpha > 0$, visible even for small
$\alpha = \pm 0.1$.
  
 \section{Linear spin-wave theory: qualitative agreement vs. failure} 
 
 To discuss the discrepancy between the numerically
 established gap on TL at $h=0$, and gapless spin waves predicted by
 LSWT, we perform LSWT calculations by mainly following Ref. \onlinecite{toth15}.
After obtaining classical ground states with three (A, B, C) or two
sublattices (A, B) on TL or SqL and HcL, respectively (see also Fig.~\ref{fig1a}), one performs
bosonisation and writes the Hamiltonian in terms of bosonic operators
($a_i$, $a_i^\dagger$).

Let us start the discussion at $h=h_c=3\alpha SJ$, i.e., at
 the magnetization plateau on TL, where the LSWT works best.
  In such
  a case, the classical angles of spins for the z-axis for each
  sublattice are $\theta_A=\pi$, $\theta_B=\theta_C=0$ which leads to a
  simplified analysis.
  Then, the exchange term between site $i$ on A
  sublattice and site $j$ on the B (or C) sublattice (for antiparallel
  spins) reads
\begin{eqnarray}
H_{J,AB}&=& -\alpha JS( a_i a_j+a_i^\dagger a_j^\dagger) +
          JS(a_i^\dagger a_i +a_j^\dagger a_j)  \nonumber\\ 
     &&     - J a_i^\dagger a_i a_j^\dagger a_j - JS^2. \label{hab}
\end{eqnarray}
Similarly, one obtains the exchange term between site $i$ on the B sublattice
and site $j$ on the C sublattice (for parallel spins)
\begin{eqnarray}
H_{J,BC}&=& \alpha JS( a_i a_j^\dagger+a_i a_j^\dagger) 
         - JS(a_i^\dagger a_i +a_j^\dagger a_j)  \nonumber\\ 
     &&     + J a_i^\dagger a_i a_j^\dagger a_j + JS^2. \label{hbc}
\end{eqnarray}
The Hamiltonian $H_{J,BC}$ contains a hopping term on the HcL, yielding an almost exact description of the one-magnon excitation (lower-branch dispersion \cite{ulaga24}) at the $m=1/3$ plateau. This term is also
governing the small number of magnon excitations, which are realized
at reduced  $m<1/3$. However, when approaching $m\sim 0$, the number of
magnons increases, and other terms become important. Among them, a
strong repulsion term $ J a_i^\dagger a_i a_j^\dagger a_j$ can play
a significant role, making the system strongly interacting/correlated.
It is, however, neglected within the standard LSWT. 

We perform standard LSW calculations on HcL, SqL and TL
to evaluate the magnetization curves  $m(h)$.
For each $h$  we first calculate the classical gs and
then include quantum corrections following
Ref.~\onlinecite{zhitomirsky98} to obtain $E_0(h)$ and from it the magnetization
$m$ as $dE_0(h)/dh$.
In contrast to the classical solution alone (which for $\alpha \ll 1$ evidently does 
not produce a proper $m$ away from the plateau regime), the obtained 
magnetization is shown in Figs.~\ref{fig1}, \ref{fig4}
and agrees qualitatively well with the full numerical calculations
of the model. It can be expected that the agreement is best close to
the magnetization saturation or plateau, where only a small number of
magnons are present in the system, therefore higher order terms, including magnon-magnon
interactions, play only a minor role. The agreement close to $m=1$ on SqL
and HcL in Fig.~\ref{fig1} is indeed very good. As one moves to
$m < 1$, or $m < 1/3$ on TL, respectively,  the discrepancy becomes larger, 
in particular for  $\alpha \ll1 $, since the neglected magnon-magnon interactions beyond 
LSWT become more relevant. On TL there is also a marked discrepancy at  
$h \sim 0$, since LSWT predicts negative $m$. This can be traced
back to the overestimate of the quantum corrections to the classical
result.
 
 \section{Discussion}
 
 In this paper, we present further results relevant for the anisotropic
 easy-axis Heisenberg model on the triangular lattice. We first show that
 some relevant features of the full model are also reproduced on
 reduced models, obtained via explicitly breaking the
 translational symmetry by fixing a third of the spins. This leads to
 the effective AHM on a HcL, but now (to keep the
 correspondence) at a finite external field $h$. Since HcL is bipartite, its behaviour is expected to be similar to the AHM on the
 SqL.  Our numerical $T=0$ results for $m(h)$ as well as for 
 the transverse magnetization
 (corresponding to the supersolid parameter on TL) $m_\perp(m)$ indeed
 reveal qualitative similarity to the full model on TL, provided that
 we are restricted to the regime of correspondence, i.e.,
 $1/2 \leq m \leq 1$, and also to stronger anisotropy
 $\alpha \ll 1$. Nevertheless, in qualitative contrast to
 AHM on TL, both HcL and SqL, even at the most interesting
 $m =1/2$, indicate finite (although reduced) $m_\perp$, i.e., confirming
gapless excitations in effective models.

\begin{figure}[t]
\centering
\includegraphics[width=0.7\columnwidth]{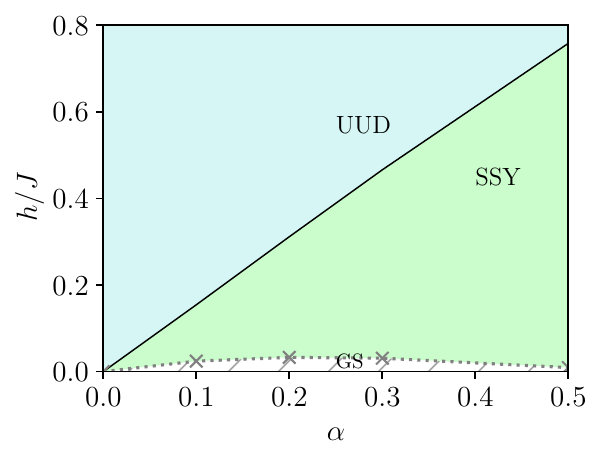}
\caption{ The ground-state phase diagram $h/J$ vs. $\alpha$ obtained in this work,
displaying three phases: the gapped spin solid (GS) phase for $h<h^*$, the 
supersolid Y (SSY) phase for $h_* < h <h_c$ and the polarized UUD 
phase on the magnetization plateau $h>h^*$. } \label{fig7}
\end{figure}

 This brings us back to the most challenging question of 
 the vanishing $m_\perp$ and also of the related magnon gap
 $\Delta_1 > 0$ in the full TL model at $h =0 $. This also opens
 the possibility of a transition/crossover at
 $\alpha < 1$, since the isotropic case $\alpha =1$ has quite well
 (even quantitatively) established $m_\perp > 0$.  We present here
 additional results for magnon gaps $h^* = \Delta_1$, as they emerge
 at $T=0$ from the analyses of $m(h)$ obtained from
 ED and DMRG results on systems up to $N = 72$ sites, covering the whole
 relevant range $\alpha < 1$, but as well some $\alpha < 0$ cases.
 Our analysis consistently reveals a decrease of the normalized gap
 $\Delta_1/ \alpha$ with increasing $\alpha < \alpha^* \sim 0.5$, as
 well as an asymmetry in $\alpha\to-\alpha$.  Similar
 conclusions follow from the new calculation of the spin stiffness $\rho_s$,
 where extrapolated $N \to \infty$ results also indicate on
 $\rho_s =0 $ (i.e., on a gapped solid) in the regime
 $\alpha < \alpha^* \sim 0.2$. The corresponding gs phase diagram, which 
 emerges from  our results (shown for $\alpha \leq 0.5$), is presented  in Fig.~\ref{fig7},
 where, in contrast to some previous studies (see e.g. \cite{yamamoto14}),
 there is (quite a narrow regime of) a gapped spin solid phase at $h< h^*(\alpha)$ 
 Still, it seems to be beyond present numerical capabilities or analytical understading to clarify whether we are 
 dealing with a transition or crossover at a particular $\alpha^*$.

The absence of supersolid in TL model at small $\alpha \ll 1$ opens an
evident question of the relation to previous results being in favor of the supersolid in the same
regime. It should be recalled that one argument  \cite{melko05,wang09,jiang09} emerges from 
the presumed symmetry of the model with respect to the sign of $\alpha$ in the Ising limit
$|\alpha | \ll 1$. In fact, an ED calculation restricted to the degenerate Ising gs 
manifold (also referred to as the dimer model \cite{wang09}) reveals upon introduction of 
finite $|\alpha| \ll 1$ that excitations are symmetric upon $\alpha\to-\alpha$.  Still, our results
(for up to $N = 36$ sites) within this restricted problem are consistent with a finite magnon gap
quite similar to the one obtained within the full model. This leads to
a possible explanation that the asymmetry and the established supersolid at
$\alpha < 0$ \cite{wessel05,heidarian05,melko05} emerge from higher
orders/terms in $\alpha$ beyond the dimer model.
    
The emergence of the gap also indicates
the qualitative failure of the LSWT in the $h \sim 0$ and
$\alpha \ll 1$ regime. Although frequently used in the interpretation
of experimental and numerical results, the LSWT has evident restrictions
in $S=1/2$ models. While starting at the finite
magnetization plateau $m=1/3$ for TL, as well as with $m=1$ for HcL
and SqL, the expansion in terms of magnons is well under control, the
neglected terms (representing magnon repulsion) become the largest
ones and apparently dominant at larger magnon densities, in
particular at $m=1/2$ for the HcL and SqL, and for $m=0$ for TL. 
The  latter case corresponds to commensurate magnon filling, which
together with strong magnon repulsion can lead to gap formation, in
analogy with the Mott-insulator mechanism. 
 
Finally, the main motivation for our study remains the recent
fascinating experiments on the KCSO material \cite{zhu24,chen24,zhu124}.
While their overall interpretation in terms of the AHM of TL
with small $\alpha \sim 0.07$ is not in question, the possible
indication of the gapped solid (at $h=0$) instead of supersolid
requires further attention. Even in the available neutron-scattering
results \cite{chen24,zhu124}, the magnon spectra at the $K$ point seem closer 
to the interpretation of a combination of quasielastic peak and gapped magnon
excitation. Moreover, 
the measured (also calculated with QMC) magnetization curves $m(h)$ \cite{zhu124}
shows the tendency of flattening at $m \sim 0$ in spite of finite $T>0$.
Still, the main indication/confirmation would be an experimental measurement of 
$m_\perp(h)$, being a reachable challenge.
  
\section*{Acknowledgments}
 
We thank A. Zheludev for a stimulating discussion of recent experimental results. 
M.U. thanks J. Hofmann for insightful discussions.
This work is supported by the Slovenian 
Research and Innovation Agency (ARIS) under program P1-0044 and project J1-50008, and the JSPS KAKENHI (Grants No. 24K00560 and No.25H01248) from the MEXT, Japan.
A part of the computational work was performed using the computational resources of the supercomputer FUGAKU provided by the
RIKEN Center for Computational Science through the HPCI System Research Project (Project ID No. hp250057). Parts of the ED calculation were performed by the XDiag library \cite{wietek2025xdiag}.

\appendix

\section{DMRG approach}
\label{dmrg}

\begin{figure}[t]
\centering
\includegraphics[width=0.6\columnwidth]{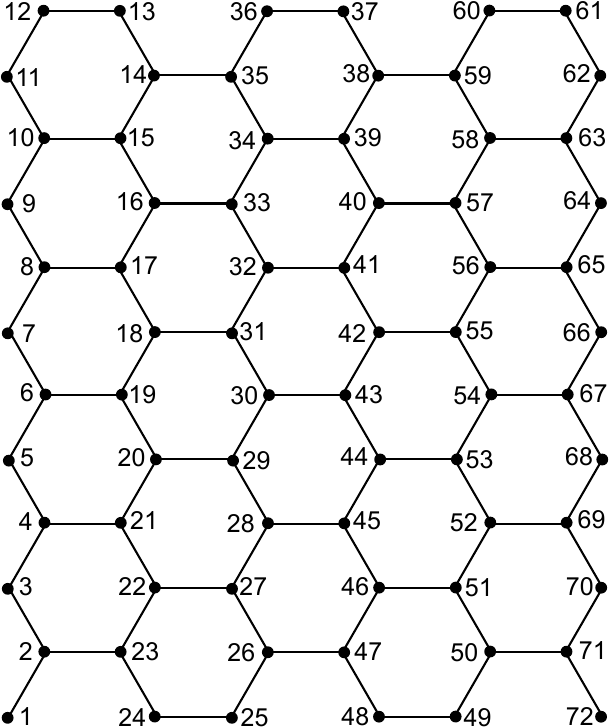}
\caption{Honeycomb lattice with $N=72$ sites. A snakelike one-dimensional chain, constructed by following the numbering assigned to each site, is employed for the DMRG calculation.}
\label{FigA1}
\end{figure}

To perform DMRG on the honeycomb lattice with $N=72$, we construct a snakelike one-dimensional chain that runs from site 1 to site 72, as shown in Fig.~\ref{FigA1}. For a square lattice with $N=8\times 8=64$, a similar one-dimensional mapping is employed (not shown). Periodic boundary conditions are imposed on both lattices. In the DMRG procedure, we set the maximum truncation number to $M=8000$. After four sweeps, the resulting truncation error is less than $1\times10^{-5}$. 

\section{Magnetization curves}
\label{chap_append}

\begin{figure*}[ht]
    \centering
    \includegraphics[width=0.8\linewidth]{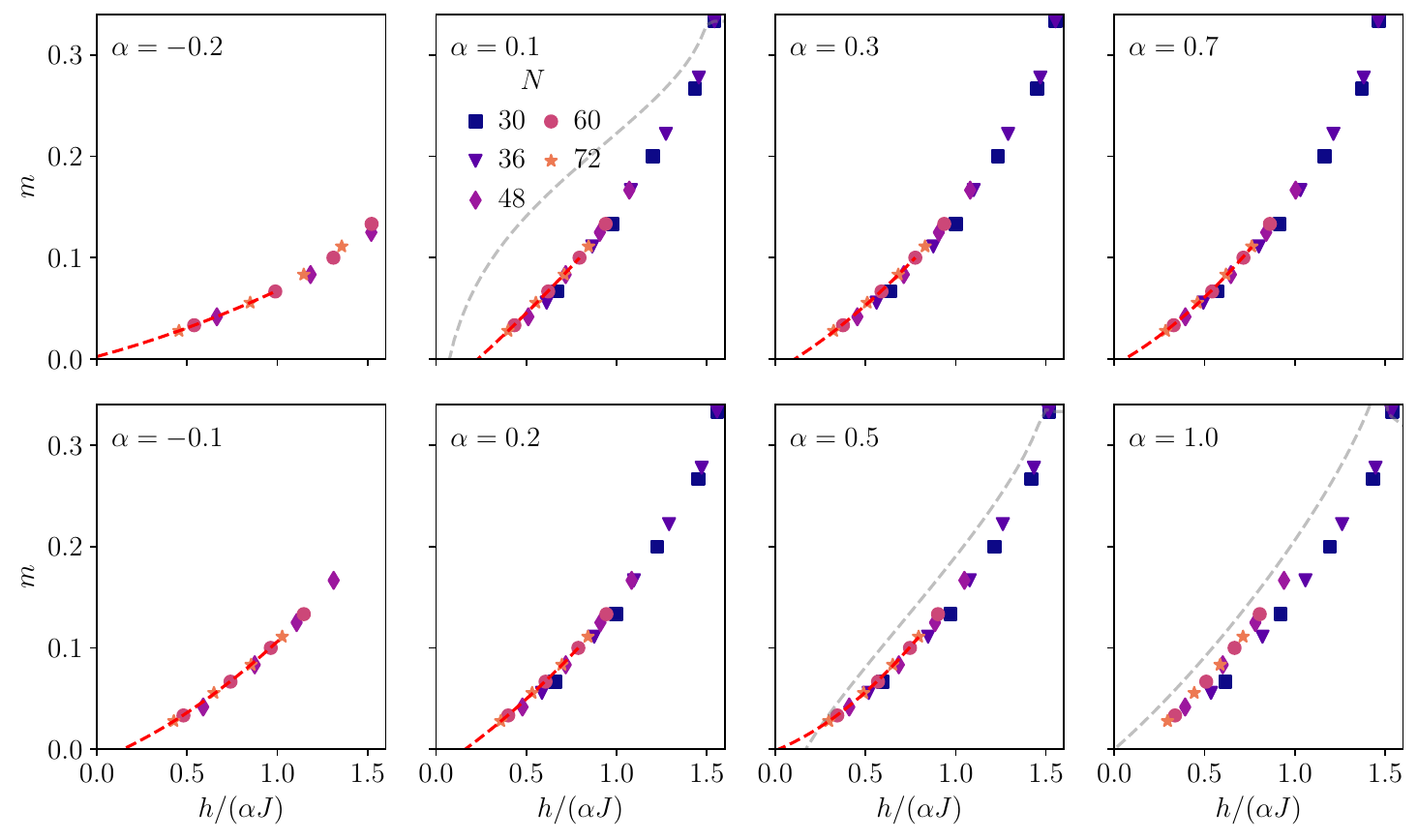}
    \caption{Magnetization curves $m$ vs. normalized fields $h/(\alpha J)$ 
    for various anisotropies $-0.2 \leq \alpha \leq 1$. 
    The thin gray lines represent LSWT results,
      while the red lines are simple polynomial fits to the data for
      all cluster sizes $N$ for small $m$. LSWT results for $\alpha=1$
    are obtained for coplanar classical spin orientations.}
    \label{fig_app}
\end{figure*}

We present a larger set of results for the magnetization curves $m(h)$,
covering the whole regime $-0.2 \leq \alpha \leq 1$ in
Fig.~\ref{fig_app}, again obtained with ED for $N \leq 36$ and
 via DMRG and $N = 48, 60, 72$ (restricted to $S^z_{\textrm{tot}} \le 4$).
 We observe that, apart from the isotropic case
$\alpha=1$, the data from different $N$ collapse quite well on a fitted curve. 
There is a notable difference between the data for positive and negative $\alpha$, i.e.,
the magnetization curves for $\alpha<0$ appear much shallower. For
$\alpha=1$, the data collapse is worse, signaling enhanced finite
size effects persisting up to $N=72$, presumably requiring also different 
scaling with $N$ for the isotropic case \cite{huse88} and  making an extrapolation $m\to 0$
less tractable.

%


\end{document}